\documentclass[twocolumn,showpacs,preprintnumbers]{revtex4}
\usepackage{amssymb}
\usepackage{amsmath}
\usepackage{graphicx}
\usepackage{epsfig}

\begin{document}

\title{Multiple 0 - $\pi$ transitions in SIFS Josephson tunnel junctions}
\author{F. Born}
\affiliation{Karlsruhe University, Institute of Micro- and Nanoelectronics Systems,
Karlsruhe, Germany}
\author{M. Siegel}
\affiliation{Karlsruhe University, Institute of Micro- and Nanoelectronics Systems,
Karlsruhe, Germany}
\author{E. K. Hollmann}
\affiliation{Research Center J\"{u}lich, ISG and IFF, J\"{u}lich, Germany}
\author{H. Braak}
\affiliation{Research Center J\"{u}lich, ISG and IFF, J\"{u}lich, Germany}
\author{A. A. Golubov}
\affiliation{Faculty of Science and Technology, University of Twente, Enschede, The
Netherlands}
\author{D. Yu. Gusakova}
\affiliation{Institute of Nuclear Physics, Moscow State University, Moscow, Russia}
\author{M. Yu. Kupriyanov}
\affiliation{Institute of Nuclear Physics, Moscow State University, Moscow, Russia}

\begin{abstract}
We report on experimental studies of superconducting coupling
through a thin Ni$_{76}$Al$_{24}$ film. A new patterning process has
been developed, which allows in combination with the wedge shaped
deposition technique the in situ deposition of 20 single
Nb/Al/Al$_{2}$O$_{3}$/Ni$_{3}$Al/Nb multilayers, each with its own
well defined Ni$_{3}$Al thickness. Every single multilayer consists
of 10 different sized Josephson junctions, showing a high
reproducibility and scaling with its junction area. Up to six damped
oscillations of the critical current density against F-layer
thickness were observed, revealing three single 0-$\pi$-transitions
in the ground state of Josephson junctions. Contrary to former
experimental studies, the exponential decay length is one magnitude
larger than the oscillation period defining decay length. The
theoretical predictions based on linearized Eilenberger equations
result in excellent agreement of theory and experimental results.
\end{abstract}

\maketitle

%

In the past few years there was a noticeable interest to the unconventional
Josephson junctions \cite{Gol04,Bu04,Ber05}, in particular, to the so-called
$\pi $ - junctions having negative critical current. These junctions provide
a $\pi $ - shift in the ground state and were realised experimentally in SFS
(superconductor-ferromagnet-superconductor) and some HTS structures.

The intensive experimental study of \textquotedblleft 0\textquotedblright\
--\textquotedblleft $\pi $\textquotedblright\ transition in SFS Josephson
junctions \cite{Ry03,Se03,Blu04,Su02,Be02,She06,Obo05,Kon02,Wei05,Se04}
confirms the existence of critical current oscillations upon the thickness
of ferromagnetic interlayer $d_{f}$. Different structure of SFS sandwiches
and SIFS tunnel junctions having been fabricated up to now. They contain
regions which are controlling the critical current and difficult to control
in experiment and describe in theory. They are SF interfaces, dead layers,
and the region in S banks with suppressed superconductivity. Contrary to
that the bulk properties of F material can be better controlled and well
described by theoretical models based on quasiclassical theory of
superconductivity. These theories predict that for large thickness of
ferromagnet the critical current of SFS junctions have to exhibit a damped
oscillations as a function of $d_{f}$
\begin{equation}
I_{c}(d_{f})=I_{c}(d_{0})\frac{\left\vert \sin (\frac{d_{f}-d_{1}}{\xi _{F2}}%
)\right\vert }{\left\vert \sin (\frac{d_{1}-d_{0}}{\xi _{F2}})\right\vert}%
\exp \left\{ -\frac{d_{f}-d_{0}}{\xi _{F1}}\right\} .  \label{eq1}
\end{equation}%
Here $d_{1}$ is the position of the first minima, $I_{c}(d_{0})$ is the
first experimental value of $I_{c}(d_{f})$. These two values take into
account the resultant action of SF interfaces and their vicinities. The
oscillations are characterized by two effective lengths. They are the decay
length $\xi _{1}$, and the length $\xi _{2,}$ which determines the period of
oscillations. In dirty limit the expressions for $\xi _{1,2}$ follow from
the Usadel equations \cite{Bu92} and have the form%
\begin{equation}
\xi _{1,2}=\sqrt{\frac{D_{f}}{\sqrt{\left( \pi T\right) ^{2}+E_{ex}^{2}}\pm
\pi T}}  \label{eq2}
\end{equation}%
where $D_{\mathit{f}}$ and $E_{\mathit{ex}}$ are the diffusive coefficient
and exchange field of ferromagnetic material, respectively. In the clean
limit one can easily get from Eilenberger equations that \cite{Eil81}%
\begin{equation}
\xi _{1}^{-1}=\xi _{0}^{-1}+\ell ^{-1},\quad \xi _{0}=\frac{v_{F}}{2\pi T}%
,\quad \xi _{2}=\xi _{H}=\frac{v_{F}}{2E_{ex}}  \label{eq3}
\end{equation}%
where $l\gg \xi _{0}$ is the electron mean free path and v$_{F}$ is the
Fermi velocity in a ferromagnet.

It is clearly seen from Eqs.~(\ref{eq2}) and ~(\ref{eq3}) that for dirty
materials $\xi _{2}>\xi _{1}$, and in the limit of large exchange energy, $%
E_{ex}>>\pi T$, the characteristic lengths are nearly equal $\xi _{1}\simeq
\xi _{2}$. In the clean limit these $\xi _{1}$ and $\xi _{2}$ are completely
independent.

Our analysis of both the bulk properties of ferromagnet materials \cite{Ku06}
and the experimental data \cite%
{Ry03,Se03,Blu04,Su02,Be02,She06,Obo05,Kon02,Wei05,Se04} has shown (see
Table~\ref{tab:tabelle1}) that in dilute ferromagnets \cite%
{Se03,Be02,Obo05,Kon02} the electron mean free path is very small providing
the fulfilment of the dirty limit conditions for the F interlayer. In these
experiments $\xi _{1}\simeq \xi _{2}$ as it follows from Eq.~(\ref{eq2}).
Contrary to that, in the structures with Ni interlayer the relation between $%
\xi _{2}$ and $\xi _{1}$ is just the opposite so that more complex model
\cite{Ber01} should be used for the data interpretation. It is necessary to
point out that in all previous experiments (except \cite{Se04}) the
structures were not fabricated in one run, so that the certain degree of
non-reproducibility of magnetic constants of F materials occurred for
junctions with different $d_{f}$. This results in increase of spread of data
with increasing $d_{f}$ and did not permit to observe the large amount of
oscillations.
\begin{table}[tbp]
\caption{Characteristic lengths in ferromagnetic materials for SFS Josephson
junctions.}
\label{tab:tabelle1}%
\begin{ruledtabular}
\begin{tabular}{cccccc}
    Ref. & $\xi _{1}$(nm) & $\xi _{2}$(nm) & F-material & $v_{F}$(m/s) & E$_{ex}$(K) \\
\hline
[8] & 1.2 & 1.6 & Fe$_{20}$Ni$_{80}$ & $2.2\cdot 10^{5}$ & 1100 \\
\lbrack 11] & 1.8 & 2 & Pd$_{0.9}$Ni$_{0.1}$ & $2\cdot 10^{5}$ & 400 \\
\lbrack 10] & 1.2 & 3.5 & Cu$_{0.53}${Ni}$_{0.47}$ &  & 850 \\
\lbrack 5] &  &  & Cu$_{0.52}$Ni$_{0.48}$ &  &  \\
\lbrack 9] & 1.7 & 1 & Ni & $2.8\cdot 10^{5}$ & 2300 \\
this work & 4.6 & 0.45 & Ni$_{3}$Al & $1.5\cdot 10^{5}$ & 1000 \\
\end{tabular}
\end{ruledtabular}
\end{table}

In this work, we improved the reproducibility of the junction parameters by
preparing the structures in one run with F - layer thickness between 10 nm
and 20 nm. For the first time, we succeeded in observation of up to six
damped oscillations of critical current with F - layer thickness. To do this
we have used the \textquotedblright wedge\textquotedblright\ shaped F -
layer technique and a new ferromagnetic material (Ni$_{3}$Al). We
experimentally obtained a one order of magnitude difference between $\xi _{1}
$ and $\xi _{2} $ which is consistent with theory based on the Eilenberger
equations \cite{Eil81}.
\begin{figure}[bp]
\includegraphics[width=3in ]{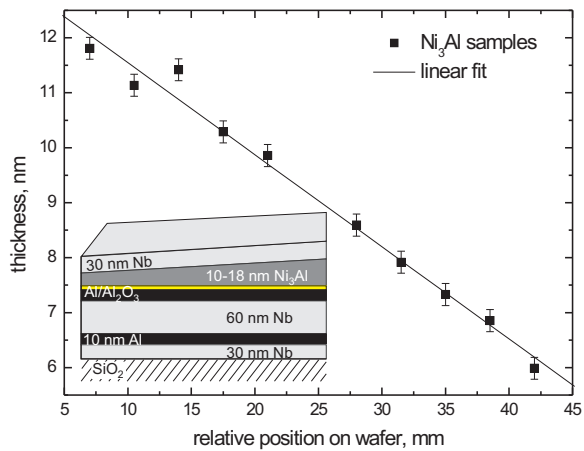}
\caption{Ni$_{76}$Al$_{24}$ film thickness measured at different positions
on a 2 inch Si wafer with the RBS method. Inset: layer sequence in a cross
section.}
\label{fig:fig1}
\end{figure}

The bottom electrode of SIFS samples consists of Nb/Al/Nb/Al and was
deposited on oxidized 2 inch Si wafers with argon magnetron sputtering. The
top 10~nm thick Al layer has been oxidized for 2 minutes in a 0,1~mbar pure
oxygen. The following Ni$_{3}$Al interlayer was sputtered with neon gas from
one single target. The target composition of the alloy was determined by
Rutherford Backscattering (RBS) and is Ni$_{74}$Al$_{26}$. A 30 nm Nb top
layer was deposited in situ to prevent the interlayer from oxidation. A
schematic cross section of the deposited multilayer is shown in insert of
Fig. 1. To achieve a thickness - gradient of the Ni$_{3}$Al layer the
position of the target above the substrate has been shifted during
deposition for several centimeter. This permits us to produce a rather
linear thickness gradient over the whole 2 inch substrate (see Fig.~\ref%
{fig:fig1}).
\begin{figure}[bp]
\includegraphics[width=3in ]{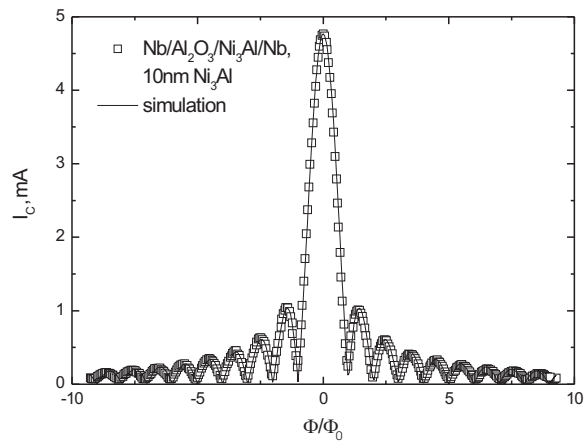}
\caption{Fraunhofer\nobreakspace-\nobreakspace pattern of a typical circular
Nb/Al/Al$_{2}$O$_{3}$/Ni$_{3}$Al/Nb Josephson junction, measured at 4.2K
(rectangles). The junction size is about 1000$\protect\mu $m$^{2}$. The
solid line results from a fit to the Fraunhofer function.}
\label{fig:fig2}
\end{figure}

\begin{figure}[tp]
\includegraphics[width=3in ]{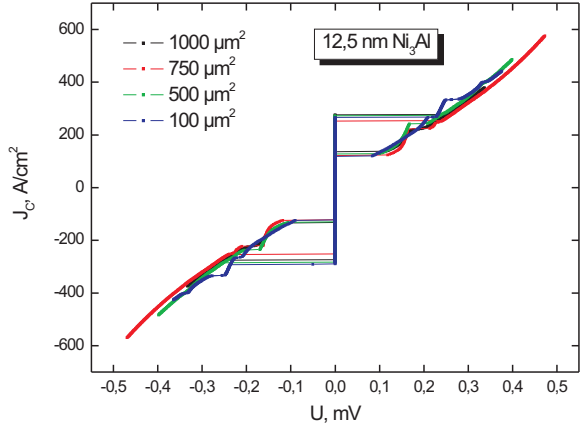}
\caption{Current density vs. voltage characteristics of four different
Josephson junctions from one patterned line with different junction sizes,
measured at 4.2 K. The size of junctions changes from 100~$\protect\mu $m$%
^{2}$ to 1000~$\protect\mu $m$^{2}$. The F-layer thickness is 12.5~nm.}
\label{fig:fig3}
\end{figure}

The transport measurements were performed at 4,2 K. Figure~\ref{fig:fig2}
shows the dependence of the critical current, $I_{C}$, upon external
magnetic field $H$. The rather optimal agreement between experimental
results and the Fraunhofer function fit indicates a uniform and homogenous
current distribution in the junction. The current-voltage characteristics
(CVC) of several Josephson junctions can be seen in Fig.~\ref{fig:fig3}.
They show a clear superconducting tunnelling behaviour with a hysteresis in
its curves. These four curves belong to one defined Ni$_{3}$Al layer
thickness of 12.5 nm. The current is normalized to the current density,
since the four junctions differ in their sizes. It can be seen in Fig.~\ref%
{fig:fig3}, that the current density differs only several percent, thus
showing a good reproducibility of our junctions. This result also indicates
that within one patterned line the F\nobreakspace-\nobreakspace layer
thickness variation could be neglected.The thickness gradient has been
characterized and proofed with RBS measurements.
The variations are several Angstroms. We developed a patterning process that
allows the creation of 20 different separated 500~$\mu $m wide lines,
distributed homogeneous along the F~-~layer thickness gradient over the
2~inch~wafer. Each lines consists of 10 different sized circular Josephson
junctions. The junction area differs from 5 to 1000~$\mu $m$^{2}$. The
variation of F~-~layer thickness inside an individual junction is negligible
small. The ferromagnetic properties of Ni$_{3}$Al films depend on neon
pressure. Details will be published elsewhere \cite{Bo06}. The magnetic
properties of the Ni$_{3}$Al layers were measured with a SQUID -
magnetometer. The Curie temperature for a 250 nm thick Ni$_{3}$Al layer is
74~K.

\begin{figure}[tp]
\includegraphics[width=3in ]{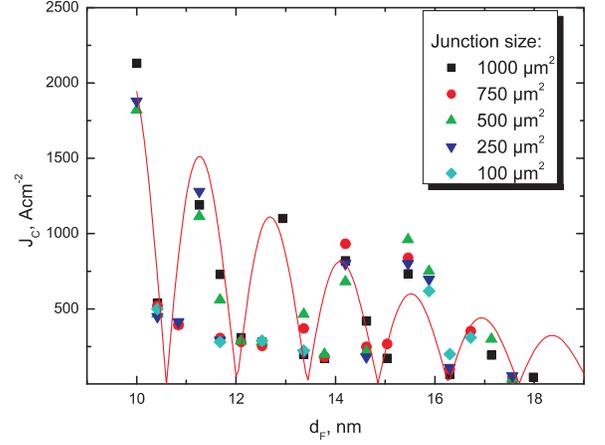}
\caption{Critical current density of up to 60 single SIFS Josephson
junctions against the F\nobreakspace-\nobreakspace layer thickness. The
solid line indicates a theoretical fit after Eq.~(\protect\ref{eq1}) with
the two fitting parameters $\protect\xi _{1}$ = 4.6 nm and $\protect\xi _{2}$
= 0.45 nm.}
\label{fig:fig4}
\end{figure}

The Ni$_{3}$Al thickness dependence of the critical current density can be
seen in Fig.~\ref{fig:fig4}. Several Josephson junctions of each line were
measured and plotted versus their corresponding Ni$_{3}$Al-layer thickness.
A clear oscillating behavior of the critical current density of more than 60
single junctions versus d$_{F}$ is shown, indicating three different 0~-~$\pi
$~transitions. The amplitude of oscillations decays exponentially with a
characteristic decay length of $\xi _{1}$=4.6~nm. It should be pointed out
that the corresponding oscillation period, given by $\pi \xi _{2}$, is one
magnitude smaller, namely $\xi _{2}$=0.45~nm. There is a rather good
agreement of the theoretical fit after Eq.~(\ref{eq1}) with these two decay
length's, see solid line in Fig.~\ref{fig:fig4}. The magnetically dead layer
has been measured for different interfaces. Corresponding to this SIFS
multilayer, we achieved a thickness of the dead layer around 5~nm to 8~nm
for each interface. So, we expected the first oscillation period at around
10~nm or even more. To find a theoretical explanation for the large
difference between $\xi _{1}$ and $\xi _{2}$ we start with linearized
Eilenberger equations which are valid at the distances from SF interface
larger than $\xi _{1}$%
\begin{equation}
2f+\frac{v_{f}\cos \theta }{\omega +iE_{ex}}\frac{d}{dx}f=\frac{\left\langle
f\right\rangle -f}{\tau \left( \omega +iE_{ex}\right) },\left\langle
()\right\rangle =\int\limits_{0}^{\pi }()\sin \theta d\theta  \label{eq4}
\end{equation}%
Here $\theta $ is the angle between direction of the Fermi velocity and
interface normal, $\tau =v_{F}/l$ is the scattering time. \newline
The solution of this equation has the form \cite{Ku81}%
\begin{equation}
f(x,\theta )=C(\theta )\exp \left\{ -\frac{x}{\xi _{eff}}\right\} ,\quad \xi
_{eff}^{-1}=\xi _{1}^{-1}+i\xi _{2}^{-1},  \label{eq5}
\end{equation}%
where $\xi _{eff}$ is the effective decay length which is independent on $%
\theta $ and $C(\theta )$ is an integration constant. Substitution of Eq.~(%
\ref{eq5}) into Eq.~(\ref{eq4}) gives
\begin{equation}
C(\theta )=\frac{\eta \left\langle C\right\rangle }{1-k^{2}\cos ^{2}\theta }%
,\;\eta =\frac{\ell ^{-1}}{\xi _{0}^{-1}+\ell ^{-1}+i\xi _{H}^{-1}},\;k=%
\frac{\eta \ell }{\xi _{eff}}  \label{eq6}
\end{equation}%
Integration of Eq.~(\ref{eq6}) over angle $\theta $ provides the equation
for $\xi_{eff}$%
\begin{equation}
\tanh \frac{\ell }{\xi _{eff}}=\frac{\xi _{eff}^{-1}}{\xi _{0}^{-1}+\ell
^{-1}+i\xi _{H}^{-1}}.  \label{eq7}
\end{equation}%
In the dirty limit, $\ell <<\xi _{0}\xi _{H}\left( \xi _{H}+\sqrt{\xi
_{0}^{2}+\xi _{H}^{2}}\right) ^{-1}$ and in the clean limit $1+\ell \xi
_{0}^{-1}>>\frac{1}{2}\max \left\{ \ln \left( 1+\ell \xi _{0}^{-1}\right)
,\ln \left( \ell \xi _{H}^{-1}\right) \right\} ,$ the solution of Eq.~(\ref%
{eq7}) reduces to Eqs.~(\ref{eq2}) and ~(\ref{eq3}), respectively. The
results of numerical solution of Eq.~(\ref{eq7}) are presented in Fig.~\ref%
{fig:fig5}.

\begin{figure}[tp]
\includegraphics[width=3in ]{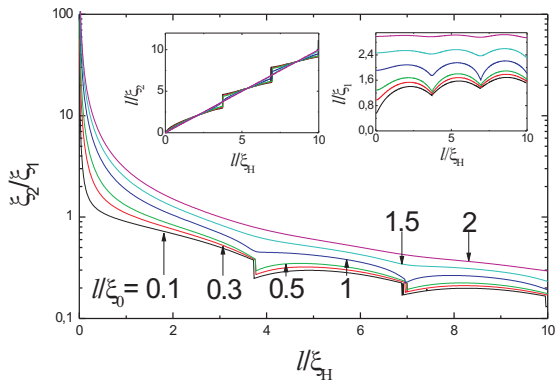}
\caption{Dependence of the ratio $\protect\xi _{2}/\protect\xi _{1}$ on
inverse magnetic length $l/\protect\xi _{H}$ calculated for different ratios
of $l/\protect\xi _{0}$. Left inset: Inverse decay length $l/\protect\xi _{2}
$ \textit{vs} inverse magnetic length $l/\protect\xi _{H}$ for different
ratios of $l/\protect\xi _{0}$. Right inset: Inverse decay length $l/\protect%
\xi _{1}$ \textit{vs} inverse magnetic length $l/\protect\xi _{H}$ for
different ratios of $l/\protect\xi _{0}$.}
\label{fig:fig5}
\end{figure}

There are steps on $\xi _{2}^{-1}(\xi _{H}^{-1})$ and $\xi _{2}/\xi _{1}$
\textit{vs} $\xi _{H}^{-1}$ dependencies (see Fig.~\ref{fig:fig5})
accompanied by the minima on $\xi _{1}^{-1}(\xi _{H}^{-1})$ curves (see Fig.~%
\ref{fig:fig5}, right inset). The ratio of $\xi _{2}/\xi _{1}$ falls very
rapidly with increase of $\xi _{H}$. It follows from Fig.~\ref{fig:fig5}
that the experimental value of this ratio $\xi_{2}/\xi _{1}\approx 0.1$ can
be achieved at $l\approx 10\xi _{H}$. For the estimated earlier parameters $%
\xi _{1}$ $\approx 4.5$ nm and $\xi _{2}$~$\approx 0.5$~nm from Fig.~\ref%
{fig:fig5}, left inset and Fig.~\ref{fig:fig5}, right inset we get nearly
equal values for the electron mean free path $l\approx 1.4\xi _{1}\approx
6,3 $nm and $l\approx 11\xi _{2}\approx 5$ nm respectively. In the last
estimation it was also supposed that $l/\xi _{0}\approx 0.1,$ resulting in $%
\xi _{0}\approx $ $50$ nm and Fermi velocity $v_{F}=1,8$\textperiodcentered $%
10^{5}$~m/s. With this value of $v_{F}$ and $\xi _{H}\approx 0.1l\approx 0.5$
nm we arrived at $E_{\mathit{ex}}\approx 1300~K\approx 0.11$~eV. This
combination of parameters is not unique. Starting from $l\approx 20\xi _{H}$
and $l/\xi _{0}\approx 0.3$ one can get $l\approx 9.5$ nm, $\xi _{0}\approx
30$ nm, $v_{F}=1.2$\textperiodcentered $10^{5}$~m/s and $E_{\mathit{ex}%
}\approx 900~K\approx 0,08$~eV. These parameters are consistent with the
previous experimental data integrated into Table~\ref{tab:tabelle1}.

The discovered behavior of $\xi _{2}$ and $\xi _{1}$ is quite general and
must be also observed in structures without ferromagnetic ordering. An
example is a normal filament of finite length, which is placed between
superconducting banks and is biased by a dc supercurrent. It was shown \cite%
{Ka06}, that the minigap induced to this filament from the S electrodes is
not a monotonous function of phase difference across the structure. This
behavior could be also explained in terms of specific dependencies of $\xi
_{2}$ and $\xi _{1}$ upon electron mean free path in current biased systems.

Summarising the presented results we conclude that utilization of the new
ferromagnetic material, Ni$_{3}$Al, as well as the wedge technique for its
deposition permits for the first time the experimental demonstration as mush
as six oscillations of the critical current as a function of thickness of
ferromagnetic layer. High reproducibility of the junctions parameters, their
scaling with the area, suppression of oscillation observed after only one
change in operation-routing sequence, namely, replacement of Ar by Ne during
the sputtering of Ni$_{3}$Al, clearly manifests that observed effect is due
to magnetic ordering in Ni$_{3}$Al film. The fact of this ordering has been
also confirmed by independent examination of magnetization of the Ni$_{3}$Al
films. It is important also to mention that Ni$_{3}$Al is an intermetallide.
This metal is widely used and well studied before \cite{Ku06}. It
successfully combines the relatively small values of exchange integral with
the transport properties close to that of strong pure ferromagnets. We
believe that it will substitute the dilute ferromagnetic alloys in the SFS
Josephson junction technology. The experimental results are consistent with
the theoretical predictions made in the frame of the Eilenberger equations.
Moreover, it was demonstrated that the intuitive knowledge about the
relation between $\xi _{2}$ and $\xi $, which is based on the dirty
theories, has a very limited field of applications and can not be used for $%
\xi _{H}$ $>5l$ or for $E_{ex}\tau >0.1$. In particular, it was for first
time recognized that the increase of $E_{ex}$ is not always accompanied by
decrease of $\xi _{1}$ sand there is some range of parameters when $\xi _{1}$
even may increase with $E_{ex}$. The fact that one may combine reasonably
large decay length with the smaller period of oscillations looks rather
attractive for possible applications of SFS Josephson junctions.

The authors thank A. D. Zaikin for useful discussions. This work was
supported in part by PI-Shift Programme, RFBR grant N$^{0}$ 06-02-90865 and
NanoNed programme under project TCS.7029.

\end{document}